\documentclass[floatfix,reprint,prl,aps,superscriptaddress]{revtex4-2}
\usepackage{times}
\usepackage{graphicx}
\usepackage{amsmath, braket, amsfonts}
\usepackage{amssymb}
\usepackage{natbib}
\usepackage{sidecap}
\usepackage{bm, color, ulem}
\usepackage{xcolor}
\usepackage{ragged2e}
\usepackage{float}
\usepackage{wrapfig,lipsum,booktabs}
\usepackage{siunitx}

\newcommand{\be}{\begin{equation}}
\newcommand{\ee}{\end{equation}}

\DeclareUnicodeCharacter{03BD}{{$\nu$}}
\DeclareUnicodeCharacter{2212}{{-}}

\makeatletter
\def\maketitle{
\@author@finish
\title@column\titleblock@produce
\suppressfloats[t]}
\makeatother

\begin{document}
\title{Hard and soft phase slips in a Fabry-P\'erot quantum Hall interferometer}
\author{N. L. Samuelson}
\thanks{These authors contributed equally to this work}
\affiliation{Department of Physics, University of California at Santa Barbara, Santa Barbara CA 93106, USA}
\author{L. A. Cohen}
\thanks{These authors contributed equally to this work}
\affiliation{Department of Physics, University of California at Santa Barbara, Santa Barbara CA 93106, USA}
\author{W. Wang}
\affiliation{Department of Physics, University of California at Santa Barbara, Santa Barbara CA 93106, USA}
\author{S.  Blanch}
\affiliation{Department of Physics, University of California at Santa Barbara, Santa Barbara CA 93106, USA}
\author{T.  Taniguchi}
\affiliation{International Center for Materials Nanoarchitectonics,
National Institute for Materials Science,  1-1 Namiki, Tsukuba 305-0044, Japan}
\author{K.  Watanabe}
\affiliation{Research Center for Functional Materials,
National Institute for Materials Science, 1-1 Namiki, Tsukuba 305-0044, Japan}
\author{M.  P. Zaletel}
\affiliation{Department of Physics, University of California, Berkeley, California 94720, USA}
\affiliation{Material Science Division, Lawrence Berkeley National Laboratory, Berkeley, California 94720, USA}
\author{A. F. Young}
\email{andrea@physics.ucsb.edu}
\affiliation{Department of Physics, University of California at Santa Barbara, Santa Barbara CA 93106, USA}
\date{\today}

%TC:ignore
\begin{abstract}
Quantum Hall Fabry-P\'erot interferometers are sensitive to the properties of bulk quasiparticles enclosed by the interferometer loop, with the interference phase containing information about both the quasiparticle statistics and the Coulomb-mediated bulk-edge coupling. Previous studies have explored the role of the bulk-edge coupling in an equilibrium picture where quasiparticles enter and exit the interferometer rapidly compared to the timescale over which the interferometer phase is measured.
Here, we present data from a monolayer graphene quantum Hall interferometer in the integer quantum Hall regime at $\nu = -1$ and $\nu = -2$.  Quantum interference shows phase slips
associated with the entrance of quasiparticles to the interferometer bulk.  Tracing the dependence of these phase slips on the magnetic field, we show that the equilibration time can become as long as several minutes. We further use  our multi-gated geometry to identify two classes of phase slips.  The first is associated with the addition of a quasiparticle to a bulk `puddle' of quasiparticles uniformly coupled to the entire chiral edge state, while the second is associated with the addition of a quasiparticle trapped by a defect site that couples predominantly to the closest portion of the edge.
\end{abstract}
\maketitle
%TC:endignore

In the quantum Hall edge-state Fabry-P\'erot interferometer, the interference signal measures the phase accumulated by the  current-carrying excitations on the edge of an interference loop. In the integer quantum Hall (IQH) regime, the measured phase arises due to the Aharonov-Bohm effect, making it proportional simply to the  magnetic flux enclosed by the edge state\cite{chamon_two_1997}:
\begin{equation}
\label{eq:theta_int}
    \frac{\theta}{2\pi} = \frac{A_I B}{\Phi_0}
\end{equation}
where $A_I$ is the interferometer area, $B$ is the applied magnetic field, and $\Phi_0 = h/e$ is the magnetic flux quantum. In the simplest picture, then, one expects a linear variation of the interference phase as the magnetic field and area (tuned by, for instance, a plunger gate voltage) are changed. However this picture is complicated by the fact that small changes in the total electron density are accommodated by the creation of localized quasiparticles in the bulk of the interference loop. These quasiparticles exert a Coulomb force on the compressible edge state, causing it to move when the number of bulk quasiparticles, $N_{qp}$, changes.
\begin{figure}[t!]
    % \centering
    \includegraphics[width = 86mm]{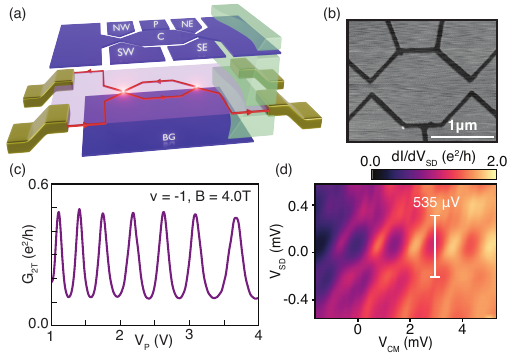}
    \caption{ \textbf{Monolayer Graphene Fabry-P\'erot interferometer.} 
    \textbf{(a)} Device schematic showing patterned  graphite top gates, hBN dielectric spacers, monolayer graphene, and graphite bottom gate. 
    \textbf{(b)} AFM topograph of the graphite top gate. 
    \textbf{(c)} High-visibility Fabry-P\'erot oscillations in the cross-device conductance as a function of the plunger gate voltage $V_P$. 
    \textbf{(d)} $dI/dV_{SD}|_{V_{CM}}$ as a function of the dc common mode voltage $V_{CM} = \frac{1}{2}(V_S + V_D)$ and source-drain difference $V_{SD}$ = $V_S - V_D$ across the sample.}
\label{fig:hs_device_details}\vspace{-20pt}
\end{figure}
Consequently, $A_I$ can have a complicated, geometry-dependent relationship to $N_{qp}$ as well as $B$, leading to discrete step-like shifts in $\theta$ as the number of enclosed quasiparticles changes.  This effect is known in the literature as bulk-edge coupling.  In this work we use the presence of these small ``slips'' in the interference phase to determine when individual quasiparticles are added to or removed from the bulk of an IQH interferometer.

\begin{figure*}[ht!]
    % \centering
    \includegraphics[width = \textwidth]{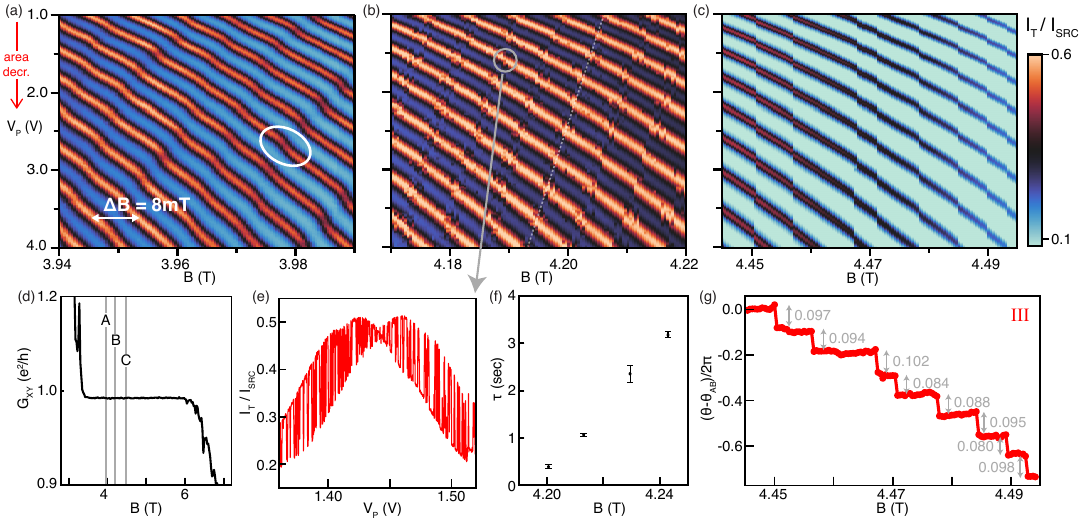}
    \caption{ \textbf{Continuous evolution of the quasiparticle charging time at $\nu=-1$.} \textbf{(a)} Transmitted current $I_T$ across the interferometer as a function of the plunger gate voltage, $V_P$, and magnetic field B, measured at a fixed electron density, with $V_B = 0.3$V and $V_C = -0.580$V within the $\nu = -1$ plateau. The interference is Aharonov-Bohm dominated throughout.  Continuous ``soft'' phase slips at $B\approx 3.96T$ evolve into 
    \textbf{(b)} ``noisy'' phase slips near $B\approx \SI{4.2}{T}$ and finally \textbf{(c)} ``hard'' phase slips for $B\approx \SI{4.47}{T}$. \textbf{(d)} Measurement of $R_{xy}$ on the W side of the device at $V_C = \SI{-0.580}{V}$ and $V_B = \SI{0.3}{V}$. The gray boxes mark the field ranges corresponding to panels a, b, and c. 
    \textbf{(e)} Trace of the transmitted current measured over 36 seconds across one of the noisy lines at the point marked in panel b.  We observed two-state switching noise corresponding to two phase-shifted interference curves. 
    \textbf{(f)} Evolution of the quasiparticle switching rate as a function of magnetic field in the ``noisy'' regime. Plotted values are the average of the state 1-to-state 2 and the state 2- to state 1 switching times $\tau \equiv \frac{1}{2}(\tau_{12}+\tau_{21})$ at the point where these values are most nearly equal, corresponding to charge degeneracy.  The error bar is the absolute difference $|\tau_{12}-\tau_{21}|$. 
    \textbf{(g)} The phase $\theta$ as a function of magnetic field, extracted from the Fourier transform of panel c. 
    The median phase difference between successive traces is subtracted to remove the smooth evolution of the Aharonov-Bohm phase. The eight discrete jumps shown in the figure have an average value of $\Delta \theta / 2\pi = -0.092$ and standard  deviation of $0.008$.}
    \label{fig:timescale}
\end{figure*}

Prior works have measured and characterized bulk-edge coupling phenomena both experimentally \cite{mcclure_fabry-perot_2012, zhang_distinct_2009, ofek_role_2010, sivan_observation_2016, nakamura_impact_2022} and theoretically \cite{rosenow2007influence, halperin_theory_2011, dinh_influence_2012}, and the role of the competing electrostatic energies in typical device geometries is now well-understood.
Often the goal is to suppress these effects as much as possible, to which end significant progress was made by engineering devices with nearby conducting layers to screen the interaction between bulk and edge; in many such devices, Aharonov-Bohm (rather than Coulomb-) dominated interference was observed in both the integer and fractional quantum Hall regimes \cite{nakamura_aharonovbohm_2019, nakamura_direct_2020, deprez_tunable_2021, ronen_aharonov-bohm_2021, nakamura_fabry-perot_2023, fu_aharonovbohm_2023,
werkmeister_strongly_2023, samuelson_anyonic_2024, werkmeister_anyon_braiding_2024, kim_aharonov-bohm_2024, kim_aharonov_bohm_even_2024}.  
These efforts have culminated in the observation of phase jumps in the fractional quantum Hall regime which can be understood as arising nearly entirely from a distinct physical mechanism --- the anyonic braiding statistics.
Notably, most of the above works have focused on the effects of bulk-edge coupling resulting from the detailed \textit{energetics} at equilibrium of the quasiparticles occupying the bulk of the interferometer. Here, we emphasize a heretofore underappreciated aspect of the physics underlying the Fabry-P\'erot interferometer, which is that the interference phase can also be used to probe the \textit{dynamics} of the quasiparticles through the time dependence of the quasiparticle occupation, $N_{qp}(t)$. While a few theoretical works have considered the effect that slow or random-in-time changes in $N_{qp}$ may have on the interference signal \cite{kane_telegraph_2003, grosfeld_switching_2006, rosenow_telegraph_2012}, experimental observations have been limited to the Coulomb-blockaded regime\cite{vandervaart_time-resolved_1994, vandervaart_time-resolved_1997}. Here we focus on the effects of a slowly time-dependent $N_{qp}$ near the Aharonov-Bohm dominated limit.

Our experimental system consists of a dual graphite-gated monolayer graphene interferometer, with a patterned top gate defining the edge trajectory as illustrated in Fig. \ref{fig:hs_device_details}a. The top graphite gate is patterned before stacking \textit{via} local anodic oxidation lithography\cite{cohen_nanoscale_2023} (see Fig. \ref{fig:hs_device_details}b) into 6 separate top gates --the plunger (P) gate, which primarily controls the interferometer area, the center (C) gate which is used to define the bulk filling factor in the cavity and on either side of the device and four additional gates  (NW, SW, NE, and SE) which are depleted in order to define  two quantum point contacts (QPCs) at which the edge states are partially back-scattered to complete the interference loop. The top- and bottom- hBN spacers are \SI{40}{nm} and \SI{50}{nm} thick, respectively, and the nominal area of the interference cavity is $0.80 \pm 0.10$ $\mu m^2$. 

Fig. \ref{fig:hs_device_details}c shows the two terminal conductance at bulk filling $\nu = -1$ as a function of $V_P$ with the exterior of the interferometer over-depleted such that $\nu_{\text{exterior}} > 0$.  The high degree of coherence is apparent from the conductance oscillations which are narrowly peaked rather than sinusoidal, indicating the presence of higher harmonics in the interference signal. The velocity of the edge modes can be extracted\cite{mcclure_edge-state_2009, chamon_two_1997} from the period of the oscillation pattern as a function of the dc source-drain voltage, $e \Delta V_{SD} = \frac{2hv}{L}$, where $v$ is the edge velocity and $L$ is the loop perimeter, and $\Delta V_{SD}$ is the period of the oscillations as a function of the source drain difference. In this measurement $V_S$ and $V_D$ are the individual voltages on the source and drain electrodes which we vary independently. Estimating $L\approx \SI{3.36}{\mu m}$ from the lithographically defined perimeter, the measured period $\Delta V_{SD} = \SI{535}{\mu V}$ gives an edge velocity $v = 2.17 \times 10^5 \SI{}{m/s}$, consistent with previous findings in graphene systems \cite{deprez_tunable_2021, ronen_aharonov-bohm_2021, fu_aharonovbohm_2023}.

Fig. \ref{fig:timescale}a shows an interference plot acquired for $\nu = -1$. 
Since $\partial A_I / \partial V_P < 0$ for $\nu = -1$, we orient \ref{fig:timescale}(a-c) to be right-handed with respect to the $B - A_I$ plane. The lines of constant phase have an negative slope consistent with ``Aharonov-Bohm'' dominated behavior, in which $A_I$ is approximately fixed as $B$ is increased at fixed $V_P$ \cite{zhang_distinct_2009}. The measured magnetic field period of \SI{8}{mT} yields an inferred area of 0.52 $\mu m^2$, slightly smaller than the nominal lithographic area. 

Notably, the negatively-sloped lines are not perfectly straight, but instead are interrupted by smooth phase slips where the phase evolves with $B$ more rapidly.  We interpret these  as corresponding to the addition or subtraction of a quasiparticle; in other words, at these points  in the $V_P - B$ plane, it becomes energetically favorable to add or expel a single quasiparticle from the interferometer. The resulting change in phase is due to the bulk-edge coupling. 

%TC:ignore
% In principle, Eq. (1) predicts that a change in $N_{QP}$ should have no effect on the interference phase in the IQHE, since $2\theta_{ex} = 2\pi$ is an unobservable phase shift. In reality, the bulk-edge coupling between the quasiparticles and the compressible edge of the interferometer allows quasiparticle addition  to change the area $A_I$ \cite{halperin_theory_2011, nakamura_direct_2020, nakamura_fabry-perot_2023}.

%The sign of the observed phase shifts in Fig.~\ref{fig:timescale}a is consistent with this effect, as after a quasiparticle exits the interferometer a reduction in $A_I$, via the application of $V_P$, is required to stay on a line of constant phase.
%TC:endignore

Figs. \ref{fig:timescale}a-c show the evolution of the phase slip behavior with increasing $B$. 
The data are acquired with  $V_P$ as the fast axis, and all other gate voltages are held constant so that increasing $B$ brings the density closer to the center of the $\nu = -1$ plateau (see Fig. \ref{fig:timescale}d). The phase slips undergo a qualitative change in behavior with increasing $B$ .  At $B\approx 4.2T$, (Fig. \ref{fig:timescale}b), phase slips are associated with  noticeable increase in the noise of the measured current near the phase slip positions.  Notably, in this regime phase slips on different constant-flux lines occur across lines of finite slope in the $V_{P}$-$B$ plane, highlighted by the white dashed line in Fig.~\ref{fig:timescale}b. 
At $B\approx 4.45T$, (Fig. \ref{fig:timescale}c), 
the noise associated with the phase slips vanishes, as does the slope of the phase slip lines in the $V_P$-$B$ plane. 
Instead, sharp discontinuities occur between successive constant-$B$ data traces. 

To quantitatively understand the evolution of the interference patterns between these regimes, we investigate the increased noise observed in Fig. \ref{fig:timescale}b. 
Fig. \ref{fig:timescale}e illustrates the conductance measured while sweeping the $V_P$ gate across one of these noisy lines at constant $B$, taken over the course of 36 seconds. The noise takes the form of two-state random telegraph noise with a few-second timescale, with the signal switching between two curves with a small phase offset. These `noisy' slips provide a natural connection between the smooth slips of Fig. \ref{fig:timescale}a and `hard' slips of Fig. \ref{fig:timescale}c.  In panel a, the quasiparticle occupation near a phase slip fluctuates faster than can be resolved in our measurement. This results in an apparently gradual phase shift, as the measured current near the charge-degeneracy point is a time-average of the phase-shifted curves corresponding to the two distinct quasiparticle occupation numbers.  
In panel b, the quasiparticle occupation remains constant over the averaging time for each point, switching only occasionally. As a result, configurations both with and without an additional quasiparticle are accessible in the same range of $V_P$. 

%TC:ignore
%\change{}{Finally, in panel c, the switching time becomes larger than time required to raster $V_P$ at constant $B$; the slips then appear vertical in the $B$-$V_P$ plane, and their location becomes irreversible and stochastic.  }
%TC:endignore

The ability  to resolve individual quasiparticle tunneling events in the time domain allows us to directly measure the characteristic time $\tau$ describing quasiparticle entry into the interferometer bulk and quantify its evolution with $B$. This is done quantitatively for the $\SI{60}{mT}$ field range presented in Fig.~\ref{fig:timescale}b.  As described in the supplementary material, points of charge degeneracy can be identified by comparing the average time spent in the $N_{qp}$ and $N_{qp}+1$ state; charge degeneracy occurs when these times are equal, with the switching time $\tau$ given by the average time between events. The quantity $\tau$ may be understood within a simple circuit model where the bulk and edge are connected by an effective resistance $R$, and $\tau = R \cdot C_{bulk}$ where $C_{bulk}  \approx \SI{1}{fF}$ is the capacitance of the interferometer bulk.
Fig. \ref{fig:timescale}f shows $\tau$ measured across a narrow range of magnetic field where $\tau$ falls in an experimentally convenient range.  $\tau$ changes over an order of magnitude very rapidly as the plateau center is approached; remarkably, the implied $R > 10^{15} \SI{}{\Omega}$ when $\tau> \SI{1}{s}$.  

The increase in $\tau$ shown in Fig. \ref{fig:timescale}f provides a natural explanation for the seemingly instantaneous nature of the phase slips in Fig. \ref{fig:timescale}c.  
When  $\tau$ becomes much larger than the $T\approx 7 s$ period of the constant-$B$ traces, the delay between tuning $V_P$ and $B$ across a charging threshold and the actual exit of a quasiparticle from the interferometer bulk will be noticeable experimentally. 
In this picture, the quasiparticles in the interferometer bulk are far from equilibrium with the edge, and the quasiparticles exit with some random delay relative to when the equilibration line is crossed, but then do not fluctuate in again before another threshold is crossed and the quasiparticle number again increases. This explains both the irregular spacing between phase slips in Fig. \ref{fig:timescale}c, and the apparent disappearance of the slope observed connecting phase slips at different values of $B$.  
Due to the sudden nature of the phase slips, their magnitude can be measured with high accuracy.  Fig. \ref{fig:timescale}g shows the evolution of the interferometer phase $\theta$ as a function of $B$ extracted from Fig. \ref{fig:timescale}c, 
after subtracting the smooth evolution due to the Aharonov-Bohm phase.  The magnitudes of the individual phase slips are marked, with $\delta \theta\in 2\pi\times(0.080 - 0.102)$ for the 8 individual slips found in the measured magnetic field range.

There are two generic possibilities concerning which  states quasiparticles occupy in the interferometer bulk.  In one scenario, the quasiparticles may enter a compressible puddle with a high density of states in the center of the interferometer in which case each added quasiparticle will contribute nearly the same degree of bulk-edge coupling as the last.  As a consequence, each individual quasiparticle is expected to produce a  nearly identical phase slip magnitude. $\tau$ in this case depends on the separation of the puddle from the edge: as the incompressible strip of quantum Hall fluid separating the puddle from the compressible edge grows (increasing the resistance across it), $\tau = R C_{bulk}$ becomes larger. 
Alternatively, each quasiparticle may be pinned by disorder to a distinct local trapping potential.  Individual quasiparticle states may then exhibit widely varying degrees of bulk-edge coupling, depending on the proximity of the pinning site to the edge. 
The data in Fig. \ref{fig:timescale}a-c are consistent with the first scenario where quasiparticles occupy a compressible puddle in the bulk, both because the phase slip magnitudes are relatively uniform and because the timescale $\tau$ increases towards the plateau center where the bulk puddle is expected to shrink.

\begin{figure}[ht!]
    \includegraphics[width = 86mm]{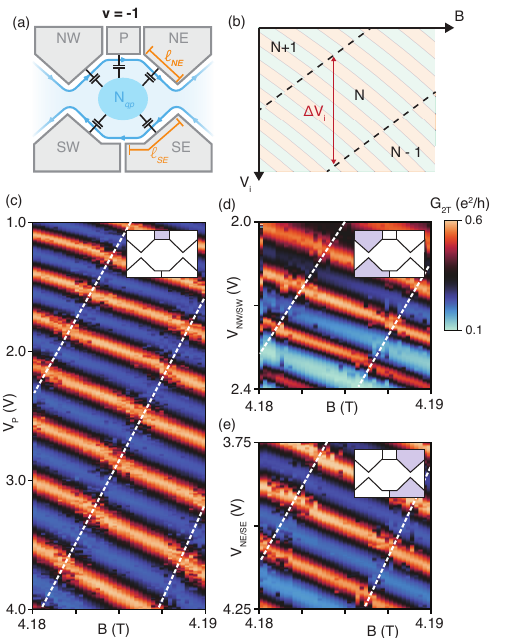}
    \caption{\textbf{Dependence of the noisy phase slip lines on the various surrounding top gates} \textbf{(a)} Schematic of the edge configuration in $\nu = -1$ with labels of the surrounding top graphite gates. If quasiparticles occupy a compressible puddle in the center, shown in blue, the capacitive coupling between the puddle and each surrounding gate per unit length should be comparable. \textbf{(b)} Schematic of the phase diagram as quasiparticle occupation of the central puddle is changed. Again note that the direction of the y-axis is oriented such that the Aharonov-Bohm lines have negative slope while the bulk is hole-doped. The noisy phase slip lines have the opposite sign of slope, and separate regions with fixed $N_{qp}$. For a set of uniformly spaced lines corresponding to charging of the central puddle, the spacing between lines gives the capacitive lever arm of the puddle to respective gate. \textbf{(c-e)} Aharonov-Bohm plots acquired in the regime of Fig. \ref{fig:timescale}  
    (b) showing noisy phase slip lines. Measurements are performed at $T=\SI{40}{mK}$, with the bulk density in $\nu = -1$. The two terminal conductance $G_{2T}$ is plotted, without subtracting the series resistances from the in-line RC filters. The y-axis (fast axis) of each plot corresponds to a different top gate(s), and the magnetic field is swept from low to high. The noisy phase slips lines, marked with dashed white lines, all show slopes with respect to all surrounding top gates in proportion with the length along which they border the perimeter of the interferometer. }
    \label{fig:qp_soup_slopes}
\end{figure}

The two scenarios can be distinguished more directly by measuring the capacitance of a given quasiparticle state to the distinct top gate electrodes. For a compressible puddle of quasiparticles, the occupation will be coupled to each top gate by a capacitance proportional to the gate length along the interferometer edge (see Fig. \ref{fig:qp_soup_slopes}a); in contrast, for a strongly localized quasiparticle state the capacitances will differ depending on the location of the trapping potential.
We test this hypothesis experimentally by measuring the interference pattern as a function of different combinations of top gate voltages in Fig. \ref{fig:qp_soup_slopes}c-e.  We focus on the `noisy' phase slip regime where phase slip positions are evident but charging events are reversible on the time scale of the gate-magnetic field scans.   The capacitance between the puddle and a given gate can be determined by the spacing of phase slips at constant $B$,  $\Delta V_i=e/C_i$. Figs. \ref{fig:qp_soup_slopes}c-e show data for, respectively, the P gate, both NW and SW gates, and both NE and SE gates.
We expect the gate-edge-to-puddle capacitance $C_i = e/\Delta V_i$ to scale linearly with the length of each gate edge $\ell_i$, which we can estimate from the AFM topography scan of the top gate as $\{\ell_{P},\ell_{NW/SW},\ell_{NE/SE}\}=\{0.47 \pm 0.07, 1.42 \pm 0.10, 1.47 \pm 0.14 \}$ $\mathrm{\mu m}$.  
From the corresponding data in Fig. \ref{fig:qp_soup_slopes}c-e, we find $\{\Delta V_P,\Delta V_{NW/SW},\Delta V_{NE/SE}\} = \{1.63\pm0.31, 0.48\pm0.12,0.71\pm0.16\}$ $\mathrm{V}$, which yields a product which is constant within uncertainty $\ell_i \times \Delta V_i=\{0.77 \pm 0.26, 0.68 \pm 0.22, 1.04 \pm 0.33\} \mathrm{\mu m\cdot V}$, consistent with a single puddle of charge capacitively coupled to the entire edge. 

The magnitude of the phase slips can be compared with the theory of bulk-edge coupling\cite{halperin_theory_2011}, which was developed for the scenario  where quasiparticles enter a compressible bulk puddle. In this theory, the phase evolution is parameterized by three quantities, $K_{IL}, K_{I}$, and $K_{L}$ where 
\begin{equation}
E = \frac{K_I}{2} (\delta n_I)^2 + \frac{K_L}{2} (\delta n_L)^2 + K_{IL} \delta n_I \delta n_L.
\end{equation}
Here $E$ is the total  energy, $\delta n_I$ is the charge on the interfering edge mode, and $\delta n_L$ is the charge in the bulk of the interferometer. 
The edge charge $\delta n_I$ is continuous and adjusts so as to minimize $E$ given $\delta n_L$ such that $\delta n_I = - \delta n_L K_{IL} / K_{I}$. 
The interference phase is proportional to the total charge contained in the interferometer\cite{halperin_theory_2011}; as a result, for each integer quasiparticle removed from or added to the bulk, the predicted magnitude of the phase shift is then $\delta \theta = \pm 2\pi K_{IL}/K_{I}$. In general, $K_{IL}$ will depend on the microscopic nature of the  state to which bulk quasiparticles are added.  For the case where quasiparticles enter a compressible puddle evenly distributed across the  bulk, an estimate of the puddle-edge electrostatic coupling when accounting for screening by the gates  gives $K_{IL} \leq \frac{1}{2} e^2 / C^g_{bulk} $, where $ C^g_{bulk} = \frac{\epsilon_z A_I}{d/2} \approx $ \SI{1.0}{fF} is the estimated geometric capacitance of the bulk puddle  (see supplementary information). The upper bound is achieved if the puddle extends to the edge; if the puddle is displaced inwards, the coupling is reduced due to screening of the puddle-edge coupling by the gates.
Experimentally, $K_I=hv/L=\SI{0.27}{meV}$ follows from the edge velocity. 
The bulk-edge coupling $K_{IL}=K_I \times \delta\theta/(2\pi)\approx 27\mu eV$ can then be obtained from the phase slip magnitudes of Fig. \ref{fig:timescale}g.  
 Together, these give $K_{IL}\approx 0.17  e^2 / C^g_{bulk}$. This value suggests a puddle of charge separated from the edge by a width of approximately $\SI{35}{nm}$ (see SI).  This separation is large compared to the magnetic length ($\ell_B \approx \SI{13}{nm}$). Tunneling across such a barrier is expected to be strongly suppressed, consistent with the long timescale $\tau$ we measure for the quasiparticle tunneling between edge and bulk.

\begin{figure}[t]
    \centering
    \includegraphics[width = 86mm]{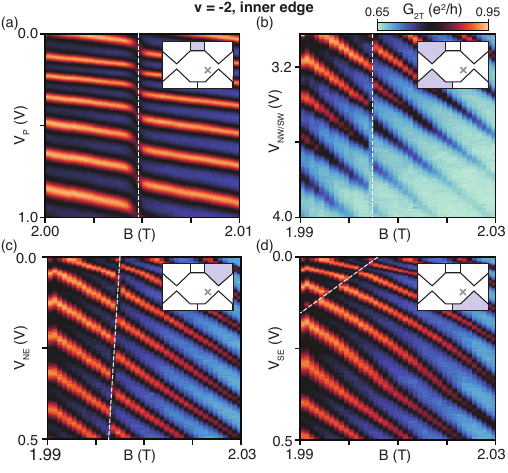}
    \caption{\textbf{Relative slopes of a soft phase slip line to each surrounding top gate} \textbf{(a-d)} Aharonov-Bohm plots acquired at $T=\SI{40}{mK}$, with the bulk density in $\nu = -2$ and the inner edge mode partially transmitted. The two terminal conductance $G_{2T}$ is plotted, without subtracting the series resistances from the in-line RC filters. The y-axis (fast axis) of each plot corresponds to a different top gate(s), with otherwise the gate voltages fixed to $V_{BG}=\SI{0}{V}$, $V_{CG}=\SI{-0.234}{V}$, $V_{NW/SW}=\SI{3.025}{V}$, $V_{NE/SE}=\SI{0}{V}$, $V_{P}=\SI{0.1}{V}$, and the magnetic field is swept from low to high. The otherwise smooth Aharonov-Bohm phase evolution (constant-phase lines having a positive slope since $\nu < 0$) is punctuated by a single reproducible phase slip at around $\SI{2.005}{T}$, marked with a dashed white line. The slope of this phase slip line in the $V$-$B$ plane depends dramatically on which gate is swept, implying the relevant quasiparticle state is located somewhere near the x marked in the insets.}
    \label{fig:pinned_qp_slopes}
\end{figure}

Interestingly, while the data acquired in Figs. \ref{fig:timescale}-\ref{fig:qp_soup_slopes} are in agreement with a picture where quasiparticles fill a compressible puddle in the bulk, we find signatures of the alternative regime proposed above (wherein added quasiparticles are pinned to local disorder sites) at a different experimental set point.  Fig. \ref{fig:pinned_qp_slopes} shows measurements at $B \approx \SI{2}{T}$ with the bulk at filling factor $\nu = -2$. Note that from the gradual occurrence of the phase slip in Fig. \ref{fig:pinned_qp_slopes}a, we infer that $\tau$ is sufficiently short that phase slip is reversible on measurement timescales.
As shown in Figs. \ref{fig:pinned_qp_slopes}a-b, the value of $B$ where the phase slip occurs is independent of $V_P$ and $V_{NW/SW}$.  In contrast, Figs. \ref{fig:pinned_qp_slopes}c-d show a weak- and strong dependence of the phase slip magnetic field on $V_{NE}$ and $V_{SE}$, respectively. 
This implies that the quasiparticle is added to a localized state located closest to the SE-gate edge of the interferometer cavity, and second-closest to the NE-gate edge, but far enough from the P, NW, and SW gates that the capacitive coupling is negligible.  This physical picture is also consistent with the large observed magnitude $\delta \theta = 2\pi\times 0.24$ of this phase slip, as might be expected for the charging of a single defect---for example, caused by a charge impurity in the hBN dielectric\cite{chiu_high_2024}---localized very near one of the sample edges. 
Although we have not studied the nature of bulk quasiparticle states systematically as a function of $\nu$ and $B$ (it is possible, for example, that a similar regime of charging localized defect states also exists at $\nu=-1$), our results suggest that which quasiparticle states are occupied involves a complex interplay between quasiparticle interactions in the bulk and the nature of disorder in van der Waals heterostructures. 

In conclusion, we have shown that Fabry-P\'erot interferometers in the quantum Hall regime can be used as a sensitive probe of quasiparticle dynamics. In particular, we have shown that phase slips are characterized by a timescale, $\tau$, related to the `RC time' to charge a given localized bulk quasiparticle state. 
Experimentally, $\tau$ can become long on typical laboratory  timescales in our graphene devices, so that the stochastic dynamics of individual charges manifest as sudden and seemingly random phase slips in our quasi-DC electronic transport measurements of the interferometer phase. We note that despite the seeming lack of reproducibility of individual charging events in this regime, stochastic behavior is precisely the expectation for a clean quantum Hall system in the low temperature limit, given the localaization of the bulk states.  In addition, the protocol we introduce to distinguish between quasiparticles which are trapped at local defect sites and those which enter a bulk compressible puddle by measuring the relative lever arms of various gates to the quasiparticle occupation may be useful for future experiments in the fractional quantum Hall states which need to reliably distinguish the effects of bulk-edge coupling and anyonic statistics.  In particular, the presence of a measurable tunneling time $\tau$ offers an intriguing possibility in the fractional quantum Hall regime; one can measure the magnitude of the observed phase-slips, in the bulk quasiparticle puddle, as a function of the tunneling time.  As the droplet shrinks, and $\tau$ increases, we expect the bulk-edge coupling to be suppressed, only leaving the contribution from the anyon phase in the interference signal for each added quasiparticle. 

A possible future improvement on this experiment would be to utilize wide bandwidth radio-frequency impedance reflectometry to greatly decrease the readout time.  In particular, for investigating the interference signatures of potential non-abelian quantum Hall states, such as the even-denominator states in bilayer graphene, this capability may be critical.  In these states the quasiparticle dynamics are likely to be considerably faster and may simply lead to complete dephasing in a quasi-DC measurement as presented in this work.  Pushing towards faster readout times will greatly improve our ability to diambiguate the effects of time-averaged quasiparticle fluctuations in the interference signal.

\clearpage
\pagebreak


\begin{thebibliography}{0}%
\makeatletter
\providecommand \@ifxundefined [1]{%
 \@ifx{#1\undefined}
}%
\providecommand \@ifnum [1]{%
 \ifnum #1\expandafter \@firstoftwo
 \else \expandafter \@secondoftwo
 \fi
}%
\providecommand \@ifx [1]{%
 \ifx #1\expandafter \@firstoftwo
 \else \expandafter \@secondoftwo
 \fi
}%
\providecommand \natexlab [1]{#1}%
\providecommand \enquote  [1]{``#1''}%
\providecommand \bibnamefont  [1]{#1}%
\providecommand \bibfnamefont [1]{#1}%
\providecommand \citenamefont [1]{#1}%
\providecommand \href@noop [0]{\@secondoftwo}%
\providecommand \href [0]{\begingroup \@sanitize@url \@href}%
\providecommand \@href[1]{\@@startlink{#1}\@@href}%
\providecommand \@@href[1]{\endgroup#1\@@endlink}%
\providecommand \@sanitize@url [0]{\catcode `\\12\catcode `\$12\catcode
  `\&12\catcode `\#12\catcode `\^12\catcode `\_12\catcode `\%12\relax}%
\providecommand \@@startlink[1]{}%
\providecommand \@@endlink[0]{}%
\providecommand \url  [0]{\begingroup\@sanitize@url \@url }%
\providecommand \@url [1]{\endgroup\@href {#1}{\urlprefix }}%
\providecommand \urlprefix  [0]{URL }%
\providecommand \Eprint [0]{\href }%
\providecommand \doibase [0]{https://doi.org/}%
\providecommand \selectlanguage [0]{\@gobble}%
\providecommand \bibinfo  [0]{\@secondoftwo}%
\providecommand \bibfield  [0]{\@secondoftwo}%
\providecommand \translation [1]{[#1]}%
\providecommand \BibitemOpen [0]{}%
\providecommand \bibitemStop [0]{}%
\providecommand \bibitemNoStop [0]{.\EOS\space}%
\providecommand \EOS [0]{\spacefactor3000\relax}%
\providecommand \BibitemShut  [1]{\csname bibitem#1\endcsname}%
\let\auto@bib@innerbib\@empty
%</preamble>
\end{thebibliography}%


\begin{thebibliography}{28}%
\makeatletter
\providecommand \@ifxundefined [1]{%
 \@ifx{#1\undefined}
}%
\providecommand \@ifnum [1]{%
 \ifnum #1\expandafter \@firstoftwo
 \else \expandafter \@secondoftwo
 \fi
}%
\providecommand \@ifx [1]{%
 \ifx #1\expandafter \@firstoftwo
 \else \expandafter \@secondoftwo
 \fi
}%
\providecommand \natexlab [1]{#1}%
\providecommand \enquote  [1]{``#1''}%
\providecommand \bibnamefont  [1]{#1}%
\providecommand \bibfnamefont [1]{#1}%
\providecommand \citenamefont [1]{#1}%
\providecommand \@href[1]{\@@startlink{#1}\@@href}%
\providecommand \@@href[1]{\endgroup#1\@@endlink}%
\providecommand \@sanitize@url [0]{\catcode `\\12\catcode `\$12\catcode `\&12\catcode `\#12\catcode `\^12\catcode `\_12\catcode `\%12\relax}%
\providecommand \@@startlink[1]{}%
\providecommand \@@endlink[0]{}%
\providecommand \@url [1]{\endgroup\@href {#1}{\urlprefix }}%
\providecommand \urlprefix  [0]{URL }%
\providecommand \Eprint [0]{}%
\providecommand \doibase [0]{http://dx.doi.org/}%
\providecommand \selectlanguage [0]{\@gobble}%
\providecommand \bibinfo  [0]{\@secondoftwo}%
\providecommand \bibfield  [0]{\@secondoftwo}%
\providecommand \translation [1]{[#1]}%
\providecommand \BibitemOpen [0]{}%
\providecommand \bibitemStop [0]{}%
\providecommand \bibitemNoStop [0]{.\EOS\space}%
\providecommand \EOS [0]{\spacefactor3000\relax}%
\providecommand \BibitemShut  [1]{\csname bibitem#1\endcsname}%
\let\auto@bib@innerbib\@empty
%</preamble>
\bibitem [{\citenamefont {Chamon}\ \emph {et~al.}(1997)\citenamefont {Chamon}, \citenamefont {Freed}, \citenamefont {Kivelson}, \citenamefont {Sondhi},\ and\ \citenamefont {Wen}}]{chamon_two_1997}%
  \BibitemOpen
  \bibfield  {author} {\bibinfo {author} {\bibfnamefont {C.~d.~C.}\ \bibnamefont {Chamon}}, \bibinfo {author} {\bibfnamefont {D.~E.}\ \bibnamefont {Freed}}, \bibinfo {author} {\bibfnamefont {S.~A.}\ \bibnamefont {Kivelson}}, \bibinfo {author} {\bibfnamefont {S.~L.}\ \bibnamefont {Sondhi}}, \ and\ \bibinfo {author} {\bibfnamefont {X.~G.}\ \bibnamefont {Wen}},\ }\href {\doibase 10.1103/PhysRevB.55.2331} {\bibfield  {journal} {\bibinfo  {journal} {Physical Review B}\ }\textbf {\bibinfo {volume} {55}},\ \bibinfo {pages} {2331} (\bibinfo {year} {1997})}\BibitemShut {NoStop}%
\bibitem [{\citenamefont {McClure}\ \emph {et~al.}(2012)\citenamefont {McClure}, \citenamefont {Chang}, \citenamefont {Marcus}, \citenamefont {Pfeiffer},\ and\ \citenamefont {West}}]{mcclure_fabry-perot_2012}%
  \BibitemOpen
  \bibfield  {author} {\bibinfo {author} {\bibfnamefont {D.~T.}\ \bibnamefont {McClure}}, \bibinfo {author} {\bibfnamefont {W.}~\bibnamefont {Chang}}, \bibinfo {author} {\bibfnamefont {C.~M.}\ \bibnamefont {Marcus}}, \bibinfo {author} {\bibfnamefont {L.~N.}\ \bibnamefont {Pfeiffer}}, \ and\ \bibinfo {author} {\bibfnamefont {K.~W.}\ \bibnamefont {West}},\ }\href {\doibase 10.1103/PhysRevLett.108.256804} {\bibfield  {journal} {\bibinfo  {journal} {Phys. Rev. Lett.}\ }\textbf {\bibinfo {volume} {108}},\ \bibinfo {pages} {256804} (\bibinfo {year} {2012})}\BibitemShut {NoStop}%
\bibitem [{\citenamefont {Zhang}\ \emph {et~al.}(2009)\citenamefont {Zhang}, \citenamefont {McClure}, \citenamefont {Levenson-Falk}, \citenamefont {Marcus}, \citenamefont {Pfeiffer},\ and\ \citenamefont {West}}]{zhang_distinct_2009}%
  \BibitemOpen
  \bibfield  {author} {\bibinfo {author} {\bibfnamefont {Y.}~\bibnamefont {Zhang}}, \bibinfo {author} {\bibfnamefont {D.~T.}\ \bibnamefont {McClure}}, \bibinfo {author} {\bibfnamefont {E.~M.}\ \bibnamefont {Levenson-Falk}}, \bibinfo {author} {\bibfnamefont {C.~M.}\ \bibnamefont {Marcus}}, \bibinfo {author} {\bibfnamefont {L.~N.}\ \bibnamefont {Pfeiffer}}, \ and\ \bibinfo {author} {\bibfnamefont {K.~W.}\ \bibnamefont {West}},\ }\href {\doibase 10.1103/PhysRevB.79.241304} {\bibfield  {journal} {\bibinfo  {journal} {Physical Review B}\ }\textbf {\bibinfo {volume} {79}},\ \bibinfo {pages} {241304} (\bibinfo {year} {2009})}\BibitemShut {NoStop}%
\bibitem [{\citenamefont {Ofek}\ \emph {et~al.}(2010)\citenamefont {Ofek}, \citenamefont {Bid}, \citenamefont {Heiblum}, \citenamefont {Stern}, \citenamefont {Umansky},\ and\ \citenamefont {Mahalu}}]{ofek_role_2010}%
  \BibitemOpen
  \bibfield  {author} {\bibinfo {author} {\bibfnamefont {N.}~\bibnamefont {Ofek}}, \bibinfo {author} {\bibfnamefont {A.}~\bibnamefont {Bid}}, \bibinfo {author} {\bibfnamefont {M.}~\bibnamefont {Heiblum}}, \bibinfo {author} {\bibfnamefont {A.}~\bibnamefont {Stern}}, \bibinfo {author} {\bibfnamefont {V.}~\bibnamefont {Umansky}}, \ and\ \bibinfo {author} {\bibfnamefont {D.}~\bibnamefont {Mahalu}},\ }\href {\doibase 10.1073/pnas.0912624107} {\bibfield  {journal} {\bibinfo  {journal} {Proceedings of the National Academy of Sciences}\ }\textbf {\bibinfo {volume} {107}},\ \bibinfo {pages} {5276} (\bibinfo {year} {2010})}\BibitemShut {NoStop}%
\bibitem [{\citenamefont {Sivan}\ \emph {et~al.}(2016)\citenamefont {Sivan}, \citenamefont {Choi}, \citenamefont {Park}, \citenamefont {Rosenblatt}, \citenamefont {Gefen}, \citenamefont {Mahalu},\ and\ \citenamefont {Umansky}}]{sivan_observation_2016}%
  \BibitemOpen
  \bibfield  {author} {\bibinfo {author} {\bibfnamefont {I.}~\bibnamefont {Sivan}}, \bibinfo {author} {\bibfnamefont {H.~K.}\ \bibnamefont {Choi}}, \bibinfo {author} {\bibfnamefont {J.}~\bibnamefont {Park}}, \bibinfo {author} {\bibfnamefont {A.}~\bibnamefont {Rosenblatt}}, \bibinfo {author} {\bibfnamefont {Y.}~\bibnamefont {Gefen}}, \bibinfo {author} {\bibfnamefont {D.}~\bibnamefont {Mahalu}}, \ and\ \bibinfo {author} {\bibfnamefont {V.}~\bibnamefont {Umansky}},\ }\href {\doibase 10.1038/ncomms12184} {\bibfield  {journal} {\bibinfo  {journal} {Nature Communications}\ }\textbf {\bibinfo {volume} {7}},\ \bibinfo {pages} {12184} (\bibinfo {year} {2016})}\BibitemShut {NoStop}%
\bibitem [{\citenamefont {Nakamura}\ \emph {et~al.}(2022)\citenamefont {Nakamura}, \citenamefont {Liang}, \citenamefont {Gardner},\ and\ \citenamefont {Manfra}}]{nakamura_impact_2022}%
  \BibitemOpen
  \bibfield  {author} {\bibinfo {author} {\bibfnamefont {J.}~\bibnamefont {Nakamura}}, \bibinfo {author} {\bibfnamefont {S.}~\bibnamefont {Liang}}, \bibinfo {author} {\bibfnamefont {G.~C.}\ \bibnamefont {Gardner}}, \ and\ \bibinfo {author} {\bibfnamefont {M.~J.}\ \bibnamefont {Manfra}},\ }\href {\doibase 10.1038/s41467-022-27958-w} {\bibfield  {journal} {\bibinfo  {journal} {Nature Communications}\ }\textbf {\bibinfo {volume} {13}},\ \bibinfo {pages} {344} (\bibinfo {year} {2022})}\BibitemShut {NoStop}%
\bibitem [{\citenamefont {Rosenow}\ and\ \citenamefont {Halperin}(2007)}]{rosenow2007influence}%
  \BibitemOpen
  \bibfield  {author} {\bibinfo {author} {\bibfnamefont {B.}~\bibnamefont {Rosenow}}\ and\ \bibinfo {author} {\bibfnamefont {B.}~\bibnamefont {Halperin}},\ }{\bibfield  {journal} {\bibinfo  {journal} {Physical review letters}\ }\textbf {\bibinfo {volume} {98}},\ \bibinfo {pages} {106801} (\bibinfo {year} {2007})}\BibitemShut {NoStop}%
\bibitem [{\citenamefont {Halperin}\ \emph {et~al.}(2011)\citenamefont {Halperin}, \citenamefont {Stern}, \citenamefont {Neder},\ and\ \citenamefont {Rosenow}}]{halperin_theory_2011}%
  \BibitemOpen
  \bibfield  {author} {\bibinfo {author} {\bibfnamefont {B.~I.}\ \bibnamefont {Halperin}}, \bibinfo {author} {\bibfnamefont {A.}~\bibnamefont {Stern}}, \bibinfo {author} {\bibfnamefont {I.}~\bibnamefont {Neder}}, \ and\ \bibinfo {author} {\bibfnamefont {B.}~\bibnamefont {Rosenow}},\ }\href {\doibase 10.1103/PhysRevB.83.155440} {\bibfield  {journal} {\bibinfo  {journal} {Physical Review B}\ }\textbf {\bibinfo {volume} {83}},\ \bibinfo {pages} {155440} (\bibinfo {year} {2011})}\BibitemShut {NoStop}%
\bibitem [{\citenamefont {Ngo~Dinh}\ and\ \citenamefont {Bagrets}(2012)}]{dinh_influence_2012}%
  \BibitemOpen
  \bibfield  {author} {\bibinfo {author} {\bibfnamefont {S.}~\bibnamefont {Ngo~Dinh}}\ and\ \bibinfo {author} {\bibfnamefont {D.~A.}\ \bibnamefont {Bagrets}},\ }\href {\doibase 10.1103/PhysRevB.85.073403} {\bibfield  {journal} {\bibinfo  {journal} {Phys. Rev. B}\ }\textbf {\bibinfo {volume} {85}},\ \bibinfo {pages} {073403} (\bibinfo {year} {2012})}\BibitemShut {NoStop}%
\bibitem [{\citenamefont {Nakamura}\ \emph {et~al.}(2019)\citenamefont {Nakamura}, \citenamefont {Fallahi}, \citenamefont {Sahasrabudhe}, \citenamefont {Rahman}, \citenamefont {Liang}, \citenamefont {Gardner},\ and\ \citenamefont {Manfra}}]{nakamura_aharonovbohm_2019}%
  \BibitemOpen
  \bibfield  {author} {\bibinfo {author} {\bibfnamefont {J.}~\bibnamefont {Nakamura}}, \bibinfo {author} {\bibfnamefont {S.}~\bibnamefont {Fallahi}}, \bibinfo {author} {\bibfnamefont {H.}~\bibnamefont {Sahasrabudhe}}, \bibinfo {author} {\bibfnamefont {R.}~\bibnamefont {Rahman}}, \bibinfo {author} {\bibfnamefont {S.}~\bibnamefont {Liang}}, \bibinfo {author} {\bibfnamefont {G.~C.}\ \bibnamefont {Gardner}}, \ and\ \bibinfo {author} {\bibfnamefont {M.~J.}\ \bibnamefont {Manfra}},\ }\href {\doibase 10.1038/s41567-019-0441-8} {\bibfield  {journal} {\bibinfo  {journal} {Nature Physics}\ }\textbf {\bibinfo {volume} {15}},\ \bibinfo {pages} {563} (\bibinfo {year} {2019})}\BibitemShut {NoStop}%
\bibitem [{\citenamefont {Nakamura}\ \emph {et~al.}(2020)\citenamefont {Nakamura}, \citenamefont {Liang}, \citenamefont {Gardner},\ and\ \citenamefont {Manfra}}]{nakamura_direct_2020}%
  \BibitemOpen
  \bibfield  {author} {\bibinfo {author} {\bibfnamefont {J.}~\bibnamefont {Nakamura}}, \bibinfo {author} {\bibfnamefont {S.}~\bibnamefont {Liang}}, \bibinfo {author} {\bibfnamefont {G.~C.}\ \bibnamefont {Gardner}}, \ and\ \bibinfo {author} {\bibfnamefont {M.~J.}\ \bibnamefont {Manfra}},\ } {\bibfield  {journal} {\bibinfo  {journal} {Nature Physics}\ }\textbf {\bibinfo {volume} {16}},\ \bibinfo {pages} {931} (\bibinfo {year} {2020})}\BibitemShut {NoStop}%
\bibitem [{\citenamefont {Déprez}\ \emph {et~al.}(2021)\citenamefont {Déprez}, \citenamefont {Veyrat}, \citenamefont {Vignaud}, \citenamefont {Nayak}, \citenamefont {Watanabe}, \citenamefont {Taniguchi}, \citenamefont {Gay}, \citenamefont {Sellier},\ and\ \citenamefont {Sacépé}}]{deprez_tunable_2021}%
  \BibitemOpen
  \bibfield  {author} {\bibinfo {author} {\bibfnamefont {C.}~\bibnamefont {Déprez}}, \bibinfo {author} {\bibfnamefont {L.}~\bibnamefont {Veyrat}}, \bibinfo {author} {\bibfnamefont {H.}~\bibnamefont {Vignaud}}, \bibinfo {author} {\bibfnamefont {G.}~\bibnamefont {Nayak}}, \bibinfo {author} {\bibfnamefont {K.}~\bibnamefont {Watanabe}}, \bibinfo {author} {\bibfnamefont {T.}~\bibnamefont {Taniguchi}}, \bibinfo {author} {\bibfnamefont {F.}~\bibnamefont {Gay}}, \bibinfo {author} {\bibfnamefont {H.}~\bibnamefont {Sellier}}, \ and\ \bibinfo {author} {\bibfnamefont {B.}~\bibnamefont {Sacépé}},\ }\href {\doibase 10.1038/s41565-021-00847-x} {\bibfield  {journal} {\bibinfo  {journal} {Nature Nanotechnology}\ }\textbf {\bibinfo {volume} {16}},\ \bibinfo {pages} {555} (\bibinfo {year} {2021})}\BibitemShut {NoStop}%
\bibitem [{\citenamefont {Ronen}\ \emph {et~al.}(2021)\citenamefont {Ronen}, \citenamefont {Werkmeister}, \citenamefont {Haie~Najafabadi}, \citenamefont {Pierce}, \citenamefont {Anderson}, \citenamefont {Shin}, \citenamefont {Lee}, \citenamefont {Lee}, \citenamefont {Johnson}, \citenamefont {Watanabe}, \citenamefont {Taniguchi}, \citenamefont {Yacoby},\ and\ \citenamefont {Kim}}]{ronen_aharonov-bohm_2021}%
  \BibitemOpen
  \bibfield  {author} {\bibinfo {author} {\bibfnamefont {Y.}~\bibnamefont {Ronen}}, \bibinfo {author} {\bibfnamefont {T.}~\bibnamefont {Werkmeister}}, \bibinfo {author} {\bibfnamefont {D.}~\bibnamefont {Haie~Najafabadi}}, \bibinfo {author} {\bibfnamefont {A.~T.}\ \bibnamefont {Pierce}}, \bibinfo {author} {\bibfnamefont {L.~E.}\ \bibnamefont {Anderson}}, \bibinfo {author} {\bibfnamefont {Y.~J.}\ \bibnamefont {Shin}}, \bibinfo {author} {\bibfnamefont {S.~Y.}\ \bibnamefont {Lee}}, \bibinfo {author} {\bibfnamefont {Y.~H.}\ \bibnamefont {Lee}}, \bibinfo {author} {\bibfnamefont {B.}~\bibnamefont {Johnson}}, \bibinfo {author} {\bibfnamefont {K.}~\bibnamefont {Watanabe}}, \bibinfo {author} {\bibfnamefont {T.}~\bibnamefont {Taniguchi}}, \bibinfo {author} {\bibfnamefont {A.}~\bibnamefont {Yacoby}}, \ and\ \bibinfo {author} {\bibfnamefont {P.}~\bibnamefont {Kim}},\ }\href {\doibase 10.1038/s41565-021-00861-z} {\bibfield  {journal} {\bibinfo  {journal} {Nature Nanotechnology}\ }\textbf {\bibinfo {volume} {16}},\ \bibinfo
  {pages} {563} (\bibinfo {year} {2021})}\BibitemShut {NoStop}%
\bibitem [{\citenamefont {Nakamura}\ \emph {et~al.}(2023)\citenamefont {Nakamura}, \citenamefont {Liang}, \citenamefont {Gardner},\ and\ \citenamefont {Manfra}}]{nakamura_fabry-perot_2023}%
  \BibitemOpen
  \bibfield  {author} {\bibinfo {author} {\bibfnamefont {J.}~\bibnamefont {Nakamura}}, \bibinfo {author} {\bibfnamefont {S.}~\bibnamefont {Liang}}, \bibinfo {author} {\bibfnamefont {G.}~\bibnamefont {Gardner}}, \ and\ \bibinfo {author} {\bibfnamefont {M.}~\bibnamefont {Manfra}},\ }\href {\doibase 10.1103/PhysRevX.13.041012} {\bibfield  {journal} {\bibinfo  {journal} {Physical Review X}\ }\textbf {\bibinfo {volume} {13}},\ \bibinfo {pages} {041012} (\bibinfo {year} {2023})}\BibitemShut {NoStop}%
\bibitem [{\citenamefont {Fu}\ \emph {et~al.}(2023)\citenamefont {Fu}, \citenamefont {Huang}, \citenamefont {Watanabe}, \citenamefont {Taniguchi}, \citenamefont {Kayyalha},\ and\ \citenamefont {Zhu}}]{fu_aharonovbohm_2023}%
  \BibitemOpen
  \bibfield  {author} {\bibinfo {author} {\bibfnamefont {H.}~\bibnamefont {Fu}}, \bibinfo {author} {\bibfnamefont {K.}~\bibnamefont {Huang}}, \bibinfo {author} {\bibfnamefont {K.}~\bibnamefont {Watanabe}}, \bibinfo {author} {\bibfnamefont {T.}~\bibnamefont {Taniguchi}}, \bibinfo {author} {\bibfnamefont {M.}~\bibnamefont {Kayyalha}}, \ and\ \bibinfo {author} {\bibfnamefont {J.}~\bibnamefont {Zhu}},\ }\href {\doibase 10.1021/acs.nanolett.2c05004} {\bibfield  {journal} {\bibinfo  {journal} {Nano Letters}\ }\textbf {\bibinfo {volume} {23}},\ \bibinfo {pages} {718} (\bibinfo {year} {2023})}\BibitemShut {NoStop}%
\bibitem [{\citenamefont {Werkmeister}\ \emph {et~al.}(2023)\citenamefont {Werkmeister}, \citenamefont {Ehrets}, \citenamefont {Ronen}, \citenamefont {Wesson}, \citenamefont {Najafabadi}, \citenamefont {Wei}, \citenamefont {Watanabe}, \citenamefont {Taniguchi}, \citenamefont {Feldman}, \citenamefont {Halperin}, \citenamefont {Yacoby},\ and\ \citenamefont {Kim}}]{werkmeister_strongly_2023}%
  \BibitemOpen
  \bibfield  {author} {\bibinfo {author} {\bibfnamefont {T.}~\bibnamefont {Werkmeister}}, \bibinfo {author} {\bibfnamefont {J.~R.}\ \bibnamefont {Ehrets}}, \bibinfo {author} {\bibfnamefont {Y.}~\bibnamefont {Ronen}}, \bibinfo {author} {\bibfnamefont {M.~E.}\ \bibnamefont {Wesson}}, \bibinfo {author} {\bibfnamefont {D.}~\bibnamefont {Najafabadi}}, \bibinfo {author} {\bibfnamefont {Z.}~\bibnamefont {Wei}}, \bibinfo {author} {\bibfnamefont {K.}~\bibnamefont {Watanabe}}, \bibinfo {author} {\bibfnamefont {T.}~\bibnamefont {Taniguchi}}, \bibinfo {author} {\bibfnamefont {D.~E.}\ \bibnamefont {Feldman}}, \bibinfo {author} {\bibfnamefont {B.~I.}\ \bibnamefont {Halperin}}, \bibinfo {author} {\bibfnamefont {A.}~\bibnamefont {Yacoby}}, \ and\ \bibinfo {author} {\bibfnamefont {P.}~\bibnamefont {Kim}},\ }\href {\doibase 10.48550/arXiv.2312.03150} {\enquote {\bibinfo {title} {Strongly coupled edge states in a graphene quantum {Hall} interferometer},}\ } (\bibinfo {year} {2023}),\ \bibinfo {note} {arXiv:2312.03150
  [cond-mat]}\BibitemShut {NoStop}%
\bibitem [{\citenamefont {Samuelson}\ \emph {et~al.}(2024)\citenamefont {Samuelson}, \citenamefont {Cohen}, \citenamefont {Wang}, \citenamefont {Blanch}, \citenamefont {Taniguchi}, \citenamefont {Watanabe}, \citenamefont {Zaletel},\ and\ \citenamefont {Young}}]{samuelson_anyonic_2024}%
  \BibitemOpen
  \bibfield  {author} {\bibinfo {author} {\bibfnamefont {N.~L.}\ \bibnamefont {Samuelson}}, \bibinfo {author} {\bibfnamefont {L.~A.}\ \bibnamefont {Cohen}}, \bibinfo {author} {\bibfnamefont {W.}~\bibnamefont {Wang}}, \bibinfo {author} {\bibfnamefont {S.}~\bibnamefont {Blanch}}, \bibinfo {author} {\bibfnamefont {T.}~\bibnamefont {Taniguchi}}, \bibinfo {author} {\bibfnamefont {K.}~\bibnamefont {Watanabe}}, \bibinfo {author} {\bibfnamefont {M.~P.}\ \bibnamefont {Zaletel}}, \ and\ \bibinfo {author} {\bibfnamefont {A.~F.}\ \bibnamefont {Young}},\ }\href {https://arxiv.org/abs/2403.19628} {\enquote {\bibinfo {title} {Anyonic statistics and slow quasiparticle dynamics in a graphene fractional quantum hall interferometer},}\ } (\bibinfo {year} {2024}),\ \Eprint {http://arxiv.org/abs/2403.19628} {arXiv:2403.19628 [cond-mat.mes-hall]} \BibitemShut {NoStop}%
\bibitem [{\citenamefont {Werkmeister}\ \emph {et~al.}(2024)\citenamefont {Werkmeister}, \citenamefont {Ehrets}, \citenamefont {Wesson}, \citenamefont {Najafabadi}, \citenamefont {Watanabe}, \citenamefont {Taniguchi}, \citenamefont {Halperin}, \citenamefont {Yacoby},\ and\ \citenamefont {Kim}}]{werkmeister_anyon_braiding_2024}%
  \BibitemOpen
  \bibfield  {author} {\bibinfo {author} {\bibfnamefont {T.}~\bibnamefont {Werkmeister}}, \bibinfo {author} {\bibfnamefont {J.~R.}\ \bibnamefont {Ehrets}}, \bibinfo {author} {\bibfnamefont {M.~E.}\ \bibnamefont {Wesson}}, \bibinfo {author} {\bibfnamefont {D.~H.}\ \bibnamefont {Najafabadi}}, \bibinfo {author} {\bibfnamefont {K.}~\bibnamefont {Watanabe}}, \bibinfo {author} {\bibfnamefont {T.}~\bibnamefont {Taniguchi}}, \bibinfo {author} {\bibfnamefont {B.~I.}\ \bibnamefont {Halperin}}, \bibinfo {author} {\bibfnamefont {A.}~\bibnamefont {Yacoby}}, \ and\ \bibinfo {author} {\bibfnamefont {P.}~\bibnamefont {Kim}},\ }\href {https://arxiv.org/abs/2403.18983} {\enquote {\bibinfo {title} {Anyon braiding and telegraph noise in a graphene interferometer},}\ } (\bibinfo {year} {2024}),\ \Eprint {http://arxiv.org/abs/2403.18983} {arXiv:2403.18983 [cond-mat.mes-hall]} \BibitemShut {NoStop}%
\bibitem [{\citenamefont {Kim}\ \emph {et~al.}(2024{\natexlab{a}})\citenamefont {Kim}, \citenamefont {Dev}, \citenamefont {Kumar}, \citenamefont {Ilin}, \citenamefont {Haug}, \citenamefont {Bhardwaj}, \citenamefont {Hong}, \citenamefont {Watanabe}, \citenamefont {Taniguchi}, \citenamefont {Stern},\ and\ \citenamefont {Ronen}}]{kim_aharonov-bohm_2024}%
  \BibitemOpen
  \bibfield  {author} {\bibinfo {author} {\bibfnamefont {J.}~\bibnamefont {Kim}}, \bibinfo {author} {\bibfnamefont {H.}~\bibnamefont {Dev}}, \bibinfo {author} {\bibfnamefont {R.}~\bibnamefont {Kumar}}, \bibinfo {author} {\bibfnamefont {A.}~\bibnamefont {Ilin}}, \bibinfo {author} {\bibfnamefont {A.}~\bibnamefont {Haug}}, \bibinfo {author} {\bibfnamefont {V.}~\bibnamefont {Bhardwaj}}, \bibinfo {author} {\bibfnamefont {C.}~\bibnamefont {Hong}}, \bibinfo {author} {\bibfnamefont {K.}~\bibnamefont {Watanabe}}, \bibinfo {author} {\bibfnamefont {T.}~\bibnamefont {Taniguchi}}, \bibinfo {author} {\bibfnamefont {A.}~\bibnamefont {Stern}}, \ and\ \bibinfo {author} {\bibfnamefont {Y.}~\bibnamefont {Ronen}},\ }\href {\doibase 10.48550/arXiv.2402.12432} {\enquote {\bibinfo {title} {Aharonov-{Bohm} interference and the evolution of phase jumps in fractional quantum {Hall} {Fabry}-{Perot} interferometers based on bi-layer graphene},}\ } (\bibinfo {year} {2024}{\natexlab{a}}),\ \bibinfo {note} {arXiv:2402.12432
  [cond-mat]}\BibitemShut {NoStop}%
\bibitem [{\citenamefont {Kim}\ \emph {et~al.}(2024{\natexlab{b}})\citenamefont {Kim}, \citenamefont {Dev}, \citenamefont {Shaer}, \citenamefont {Kumar}, \citenamefont {Ilin}, \citenamefont {Haug}, \citenamefont {Iskoz}, \citenamefont {Watanabe}, \citenamefont {Taniguchi}, \citenamefont {Mross}, \citenamefont {Stern},\ and\ \citenamefont {Ronen}}]{kim_aharonov_bohm_even_2024}%
  \BibitemOpen
  \bibfield  {author} {\bibinfo {author} {\bibfnamefont {J.}~\bibnamefont {Kim}}, \bibinfo {author} {\bibfnamefont {H.}~\bibnamefont {Dev}}, \bibinfo {author} {\bibfnamefont {A.}~\bibnamefont {Shaer}}, \bibinfo {author} {\bibfnamefont {R.}~\bibnamefont {Kumar}}, \bibinfo {author} {\bibfnamefont {A.}~\bibnamefont {Ilin}}, \bibinfo {author} {\bibfnamefont {A.}~\bibnamefont {Haug}}, \bibinfo {author} {\bibfnamefont {S.}~\bibnamefont {Iskoz}}, \bibinfo {author} {\bibfnamefont {K.}~\bibnamefont {Watanabe}}, \bibinfo {author} {\bibfnamefont {T.}~\bibnamefont {Taniguchi}}, \bibinfo {author} {\bibfnamefont {D.~F.}\ \bibnamefont {Mross}}, \bibinfo {author} {\bibfnamefont {A.}~\bibnamefont {Stern}}, \ and\ \bibinfo {author} {\bibfnamefont {Y.}~\bibnamefont {Ronen}},\ }\href {https://arxiv.org/abs/2412.19886} {\enquote {\bibinfo {title} {Aharonov-bohm interference in even-denominator fractional quantum hall states},}\ } (\bibinfo {year} {2024}{\natexlab{b}}),\ \Eprint {http://arxiv.org/abs/2412.19886} {arXiv:2412.19886
  [cond-mat.mes-hall]} \BibitemShut {NoStop}%
\bibitem [{\citenamefont {Kane}(2003)}]{kane_telegraph_2003}%
  \BibitemOpen
  \bibfield  {author} {\bibinfo {author} {\bibfnamefont {C.~L.}\ \bibnamefont {Kane}},\ }\href {\doibase 10.1103/PhysRevLett.90.226802} {\bibfield  {journal} {\bibinfo  {journal} {Phys. Rev. Lett.}\ }\textbf {\bibinfo {volume} {90}},\ \bibinfo {pages} {226802} (\bibinfo {year} {2003})}\BibitemShut {NoStop}%
\bibitem [{\citenamefont {Grosfeld}\ \emph {et~al.}(2006)\citenamefont {Grosfeld}, \citenamefont {Simon},\ and\ \citenamefont {Stern}}]{grosfeld_switching_2006}%
  \BibitemOpen
  \bibfield  {author} {\bibinfo {author} {\bibfnamefont {E.}~\bibnamefont {Grosfeld}}, \bibinfo {author} {\bibfnamefont {S.~H.}\ \bibnamefont {Simon}}, \ and\ \bibinfo {author} {\bibfnamefont {A.}~\bibnamefont {Stern}},\ }\href {\doibase 10.1103/PhysRevLett.96.226803} {\bibfield  {journal} {\bibinfo  {journal} {Phys. Rev. Lett.}\ }\textbf {\bibinfo {volume} {96}},\ \bibinfo {pages} {226803} (\bibinfo {year} {2006})}\BibitemShut {NoStop}%
\bibitem [{\citenamefont {Rosenow}\ and\ \citenamefont {Simon}(2012)}]{rosenow_telegraph_2012}%
  \BibitemOpen
  \bibfield  {author} {\bibinfo {author} {\bibfnamefont {B.}~\bibnamefont {Rosenow}}\ and\ \bibinfo {author} {\bibfnamefont {S.~H.}\ \bibnamefont {Simon}},\ }\href {\doibase 10.1103/PhysRevB.85.201302} {\bibfield  {journal} {\bibinfo  {journal} {Phys. Rev. B}\ }\textbf {\bibinfo {volume} {85}},\ \bibinfo {pages} {201302} (\bibinfo {year} {2012})}\BibitemShut {NoStop}%
\bibitem [{\citenamefont {van~der Vaart}\ \emph {et~al.}(1994)\citenamefont {van~der Vaart}, \citenamefont {de~Ruyter~van Steveninck}, \citenamefont {Kouwenhoven}, \citenamefont {Johnson}, \citenamefont {Nazarov}, \citenamefont {Harmans},\ and\ \citenamefont {Foxon}}]{vandervaart_time-resolved_1994}%
  \BibitemOpen
  \bibfield  {author} {\bibinfo {author} {\bibfnamefont {N.~C.}\ \bibnamefont {van~der Vaart}}, \bibinfo {author} {\bibfnamefont {M.~P.}\ \bibnamefont {de~Ruyter~van Steveninck}}, \bibinfo {author} {\bibfnamefont {L.~P.}\ \bibnamefont {Kouwenhoven}}, \bibinfo {author} {\bibfnamefont {A.~T.}\ \bibnamefont {Johnson}}, \bibinfo {author} {\bibfnamefont {Y.~V.}\ \bibnamefont {Nazarov}}, \bibinfo {author} {\bibfnamefont {C.~J. P.~M.}\ \bibnamefont {Harmans}}, \ and\ \bibinfo {author} {\bibfnamefont {C.~T.}\ \bibnamefont {Foxon}},\ }\href {\doibase 10.1103/PhysRevLett.73.320} {\bibfield  {journal} {\bibinfo  {journal} {Phys. Rev. Lett.}\ }\textbf {\bibinfo {volume} {73}},\ \bibinfo {pages} {320} (\bibinfo {year} {1994})}\BibitemShut {NoStop}%
\bibitem [{\citenamefont {van~der Vaart}\ \emph {et~al.}(1997)\citenamefont {van~der Vaart}, \citenamefont {Kouwenhoven}, \citenamefont {de~Ruyter~van Steveninck}, \citenamefont {Nazarov}, \citenamefont {Harmans},\ and\ \citenamefont {Foxon}}]{vandervaart_time-resolved_1997}%
  \BibitemOpen
  \bibfield  {author} {\bibinfo {author} {\bibfnamefont {N.~C.}\ \bibnamefont {van~der Vaart}}, \bibinfo {author} {\bibfnamefont {L.~P.}\ \bibnamefont {Kouwenhoven}}, \bibinfo {author} {\bibfnamefont {M.~P.}\ \bibnamefont {de~Ruyter~van Steveninck}}, \bibinfo {author} {\bibfnamefont {Y.~V.}\ \bibnamefont {Nazarov}}, \bibinfo {author} {\bibfnamefont {C.~J. P.~M.}\ \bibnamefont {Harmans}}, \ and\ \bibinfo {author} {\bibfnamefont {C.~T.}\ \bibnamefont {Foxon}},\ }\href {\doibase 10.1103/PhysRevB.55.9746} {\bibfield  {journal} {\bibinfo  {journal} {Phys. Rev. B}\ }\textbf {\bibinfo {volume} {55}},\ \bibinfo {pages} {9746} (\bibinfo {year} {1997})}\BibitemShut {NoStop}%
\bibitem [{\citenamefont {Cohen}\ \emph {et~al.}(2023)\citenamefont {Cohen}, \citenamefont {Samuelson}, \citenamefont {Wang}, \citenamefont {Klocke}, \citenamefont {Reeves}, \citenamefont {Taniguchi}, \citenamefont {Watanabe}, \citenamefont {Vijay}, \citenamefont {Zaletel},\ and\ \citenamefont {Young}}]{cohen_nanoscale_2023}%
  \BibitemOpen
  \bibfield  {author} {\bibinfo {author} {\bibfnamefont {L.~A.}\ \bibnamefont {Cohen}}, \bibinfo {author} {\bibfnamefont {N.~L.}\ \bibnamefont {Samuelson}}, \bibinfo {author} {\bibfnamefont {T.}~\bibnamefont {Wang}}, \bibinfo {author} {\bibfnamefont {K.}~\bibnamefont {Klocke}}, \bibinfo {author} {\bibfnamefont {C.~C.}\ \bibnamefont {Reeves}}, \bibinfo {author} {\bibfnamefont {T.}~\bibnamefont {Taniguchi}}, \bibinfo {author} {\bibfnamefont {K.}~\bibnamefont {Watanabe}}, \bibinfo {author} {\bibfnamefont {S.}~\bibnamefont {Vijay}}, \bibinfo {author} {\bibfnamefont {M.~P.}\ \bibnamefont {Zaletel}}, \ and\ \bibinfo {author} {\bibfnamefont {A.~F.}\ \bibnamefont {Young}},\ }\href {\doibase 10.1038/s41567-023-02114-3} {\bibfield  {journal} {\bibinfo  {journal} {Nature Physics}\ }\textbf {\bibinfo {volume} {19}},\ \bibinfo {pages} {1502} (\bibinfo {year} {2023})}\BibitemShut {NoStop}%
\bibitem [{\citenamefont {McClure}\ \emph {et~al.}(2009)\citenamefont {McClure}, \citenamefont {Zhang}, \citenamefont {Rosenow}, \citenamefont {Levenson-Falk}, \citenamefont {Marcus}, \citenamefont {Pfeiffer},\ and\ \citenamefont {West}}]{mcclure_edge-state_2009}%
  \BibitemOpen
  \bibfield  {author} {\bibinfo {author} {\bibfnamefont {D.~T.}\ \bibnamefont {McClure}}, \bibinfo {author} {\bibfnamefont {Y.}~\bibnamefont {Zhang}}, \bibinfo {author} {\bibfnamefont {B.}~\bibnamefont {Rosenow}}, \bibinfo {author} {\bibfnamefont {E.~M.}\ \bibnamefont {Levenson-Falk}}, \bibinfo {author} {\bibfnamefont {C.~M.}\ \bibnamefont {Marcus}}, \bibinfo {author} {\bibfnamefont {L.~N.}\ \bibnamefont {Pfeiffer}}, \ and\ \bibinfo {author} {\bibfnamefont {K.~W.}\ \bibnamefont {West}},\ }\href {\doibase 10.1103/PhysRevLett.103.206806} {\bibfield  {journal} {\bibinfo  {journal} {Physical Review Letters}\ }\textbf {\bibinfo {volume} {103}},\ \bibinfo {pages} {206806} (\bibinfo {year} {2009})}\BibitemShut {NoStop}%
\bibitem [{\citenamefont {Chiu}\ \emph {et~al.}(2024)\citenamefont {Chiu}, \citenamefont {Wang}, \citenamefont {Fan}, \citenamefont {Watanabe}, \citenamefont {Taniguchi}, \citenamefont {Liu}, \citenamefont {Zaletel},\ and\ \citenamefont {Yazdani}}]{chiu_high_2024}%
  \BibitemOpen
  \bibfield  {author} {\bibinfo {author} {\bibfnamefont {C.-L.}\ \bibnamefont {Chiu}}, \bibinfo {author} {\bibfnamefont {T.}~\bibnamefont {Wang}}, \bibinfo {author} {\bibfnamefont {R.}~\bibnamefont {Fan}}, \bibinfo {author} {\bibfnamefont {K.}~\bibnamefont {Watanabe}}, \bibinfo {author} {\bibfnamefont {T.}~\bibnamefont {Taniguchi}}, \bibinfo {author} {\bibfnamefont {X.}~\bibnamefont {Liu}}, \bibinfo {author} {\bibfnamefont {M.~P.}\ \bibnamefont {Zaletel}}, \ and\ \bibinfo {author} {\bibfnamefont {A.}~\bibnamefont {Yazdani}},\ }\href {https://arxiv.org/abs/2410.10961} {\enquote {\bibinfo {title} {High spatial resolution charge sensing of quantum hall states},}\ } (\bibinfo {year} {2024}),\ \Eprint {http://arxiv.org/abs/2410.10961} {arXiv:2410.10961 [cond-mat.mes-hall]} \BibitemShut {NoStop}%
\end{thebibliography}
\end{document}

\title{Supplemental Information}
\author{N. L. Samuelson}
\thanks{These authors contributed equally to this work}
\affiliation{Department of Physics, University of California at Santa Barbara, Santa Barbara CA 93106, USA}
\author{L. A. Cohen}
\thanks{These authors contributed equally to this work}
\affiliation{Department of Physics, University of California at Santa Barbara, Santa Barbara CA 93106, USA}
\author{W. Wang}
\affiliation{Department of Physics, University of California at Santa Barbara, Santa Barbara CA 93106, USA}
\author{S.  Blanch}
\affiliation{Department of Physics, University of California at Santa Barbara, Santa Barbara CA 93106, USA}
\author{T.  Taniguchi}
\affiliation{International Center for Materials Nanoarchitectonics,
National Institute for Materials Science,  1-1 Namiki, Tsukuba 305-0044, Japan}
\author{K.  Watanabe}
\affiliation{Research Center for Functional Materials,
National Institute for Materials Science, 1-1 Namiki, Tsukuba 305-0044, Japan}
\author{M.  P. Zaletel}
\affiliation{Department of Physics, University of California, Berkeley, California 94720, USA}
\affiliation{Material Science Division, Lawrence Berkeley National Laboratory, Berkeley, California 94720, USA}
\author{A. F. Young}
\email{andrea@physics.ucsb.edu}
\affiliation{Department of Physics, University of California at Santa Barbara, Santa Barbara CA 93106, USA}
\date{\today}
\maketitle
\onecolumngrid
\renewcommand\thefigure{S\arabic{figure}}
\setcounter{figure}{0}
\setcounter{equation}{0}

\section{Sample Fabrication}
The van der Waals stack was fabricated using the van der Waals dry-transfer process with a polycarbonate film on a PDMS dome. Before stacking, electrode-free AFM anodic oxidation lithography is performed on the top graphite to create the interferometer gate structure \cite{cohen_nanoscale_2023}. After stacking, the device is detached onto a conductively-doped Si substrate with \SI{285}{nm} of thermal SiO$_2$ and the polycarbonate is dissolved in chloroform.  An aluminum etch mask is defined via E-Beam Lithography (EBL) of a PMMA A4 495K/A2 950K bilayer resist before e-beam evaporation of \SI{40}{nm} of aluminum. The stack is etched using a CHF$_3$/O$_2$ plasma reactive ion etch for several minutes. The aluminum mask is removed by etching in a $<$3\% TMAH solution (AZ300MIF) for 20 minutes. Then, a contact electrode pattern is defined using a PMMA 950K A8 mask and another EBL exposure, a 30 second CHF$_3$/O$_2$ RIE plasma etch is performed to clean the exposed device edges, and edge contacts are deposited by e-Beam evaporation of 5/15/150nm of Cr/Pd/Au.

\begin{figure}[hb!]
    \centering
    \includegraphics[width = 105mm]{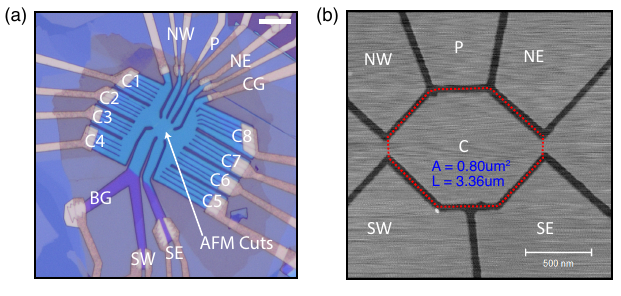}    \caption{\textbf{Device Image and AFM Cut Gate} \textbf{(a)} Optical micrograph of the device. Edge contacts to the monolayer are labeled C1-8.  Contacts to the gates are marked NW, SW, NE, SE, P, CG, and BG. Scale bar (upper right) is $\SI{10}{\mu m}$. \textbf{(b)} AFM topograph of the top graphite gate after pick-up with an hBN flake (mid-stack).  The lithographic area, defined by the perimeter drawn in the red dashed line, is $0.80 \pm 0.10\SI{}{\mu m^2}$, and the perimeter is $3.36 \pm 0.31\SI{}{\mu m}$. 
    $C_{bulk} \equiv \frac{2\epsilon_0\epsilon_\perp A}{\overline{d}} \approx \SI{1.0}{fF}$, the geometric capacitance of the interferometer bulk, can be estimated from these parameters, with $\overline{d}=\SI{45}{nm}$ being average of the top- and bottom- gate dielectric thicknesses of 40 and 50 nm respectively. The perpendicular dielectric constant of hBN is taken as $\epsilon_\perp\approx 3.25$. }
    \label{fig:device_image}
\end{figure}
\newpage 

\section{Measurement Parameters}

Experiments were performed in a dry dilution refrigerator at a base temperature of \SI{55}{mK}. Electronic RC filters on all transport lines are used to lower the electron temperature. Transport was measured at \SI{17.7777}{Hz} with SR860 lock-in amplifiers. A Basel Precision Instruments SP983c high stability I to V converter (IF3602) is used to amplify the current signals, while the voltages are measured directly with the SR860 using no additional pre-amplification.

\subsection{Fixed Gate Voltages and Magnetic Fields for Main Text Figures}
\begin{figure}[ht]
    \begin{center}
    \begin{tabular}{||c c c c c c c||} 
     \hline
     \textbf{Main Text Figure} & $V_{C}$ & $V_{BG} $ & $V_{NW/SW}$ & $V_{NE/SE}$ & $V_P$ & $B$ \\ [0.5ex] 
     \hline\hline
     Fig.~1c & -0.580V & 0.300V & 2.20V & 4.00V & - & 4.0T\\ 
     \hline
     Fig.~1d & -0.580V & 0.300V & 2.20V & 4.00V & 2.20V & 4.0T\\ 
     \hline
     Fig.~2a-b & -0.580V & 0.300V & 2.20V & 4.00V & - & - \\ 
     \hline
     Fig.~2c (g) & -0.580V & 0.300V & 2.50V & 4.00V & - & - \\ 
     \hline
     Fig.~2d & -0.580V & 0.300V & 0.0V & 0.0V & 0.0V & - \\ 
     \hline
     Fig.~3c-e & -0.580V & 0.300V & 2.20V & 4.00V & 2.00V & - \\ 
     \hline
     Fig.~4a-d & -0.234V & 0.0V & 3.025V & 0.0 & 0.1 & - \\ 
     \hline
    \end{tabular}
    \end{center}
    \caption{\textbf{Fixed voltage set points and magnetic fields for main text figures} For figures 3 and 4, the fixed voltages specified apply for each panel in which the gate is not varied as the fast axis for the measurement.}
    \label{fig:enter-label}
\end{figure}

The two terminal conductance across the device, $G_{2T}$, is measured in figures 1c, 3 and 4, with an ac voltage bias of $\SI{25}{\mu V}$, $\SI{25}{\mu V}$, and $\SI{10}{\mu V}$, respectively, \textit{without} subtracting the in-line series resistance of $\SI{23.4}{k\Omega}$ from the RC filters.

A two-ground configuration is employed for the measurements of the transmission coefficient, $I_T/I_{SRC}$, in Fig.~2: A \SI{0.9}{nA} ac current bias is applied at contact C6, and the contact C8 is grounded to sink the current reflected by the interferometer. The transmitted current is measured on the opposite side of the device, on contact C4. A constant \SI{-12}{V} is applied to the Si substrate throughout these measurements to maintain hole-doping of the contact regions.

\section{Analysis of the Source-Drain and Common-Mode Voltage Dependence}

\begin{figure}[hb!]
    \centering
    \includegraphics[width = 135mm]{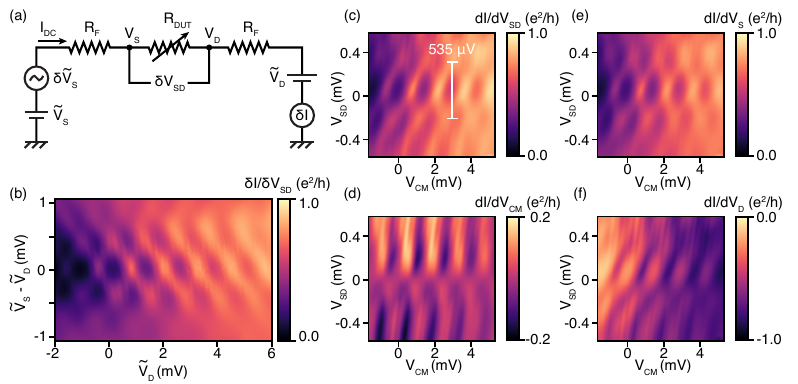}    \caption{\textbf{Source-drain and common-mode DC voltage dependence of the interference in $\nu = -1$} \textbf{(a)} Equivalent circuit model for the measurement of the DC voltage dependence of the interference \textbf{(b)} The four-terminal conductance ignoring the contribution of the filters, as a function of the \textit{applied} DC voltages $\tilde{V}_{D}$ and $\tilde{V}_{S}-\tilde{V}_{D}$. \textbf{(c-f)} $dI/dV_{SD}$, $dI/dV_{CM}$, $dI/dV_{S}$, and $dI/dV_{D}$ numerically calculated from the DC current, as a function of the corrected dc voltages $V_{SD} = V_S - V_D$ and $V_{CM}=\frac{1}{2}(V_S+V_D)$, accounting for the finite filter resistance. The period in $V_{SD}$ (marked) is $\SI{535}{\mu V}$.}
    \label{fig:vsd_analysis}
\end{figure}

A schematic for the measurement circuit is shown in Fig. \ref{fig:vsd_analysis}a. Each of the RC filters on the source and drain contacts provides a resistance of $R_F = 11.7 k\Omega$ (the capacitors are not illustrated). A fixed ac voltage bias of $\delta \tilde{V}_{S} = \SI{138}{\mu V}$ is applied, in addition to varying DC source and drain voltages $\tilde{V}_S$ and $\tilde{V}_D$. The AC current into the drain, $\delta I$ is measured, along with the diagonal voltage drop $\delta V_{SD}$. The sample resistance $R_{DUT}$ is unknown and varies with the (unmeasured) voltages $V_{SD} = V_S - V_D$ and $V_{CM} = \frac{1}{2}(V_S + V_D)$.

The finite resistance of the in-line RC filters on each transport contact leads to an actual voltage drop, $V_{SD} = V_S - V_D$ across the sample which is significantly less than the applied DC voltage, $\tilde{V}_S - \tilde{V}_D$. The DC current and voltage are not directly measured --- instead, we calculate them by integrating the measured ac voltage as a function of the applied $\tilde{V}_S$:

$$
V_{SD} = \int_{\tilde{V}_D}^{\tilde{V}_S - \tilde{V}_D} \frac{\delta V_{SD}}{\delta \tilde{V}_{S}} d(\tilde{V}_S^\prime - \tilde{V}_{D}^\prime)
$$

The DC current and DC common mode voltage can be obtained straightforwardly from $V_{SD}$:

$$
I = \frac{\tilde{V}_S - \tilde{V}_D - V_{SD}}{2R_F}
$$

$$
V_{CM}=\frac{1}{2}V_{SD}+\tilde{V}_D+IR_F
$$

We can then take numerical derivatives of the calculated DC current in any chosen direction. These derivatives are shown in Fig. \ref{fig:vsd_analysis}c-f.

The period of the conductance $dI/dV_{SD}$ is used to determine the edge velocity by its relation to the dynamical phase accrued by particles injected at a finite energy: $e\Delta V_{SD} = \frac{2hv}{L}$, where $L$ is the perimeter of the interferometer. The common-mode period differs from this --- we note that in the measured regime (the same regime as in Fig. \ref{fig:timescale}a, near a large number of soft phase slips), we expect the bulk to have a non-negligible compressibility, which means that the possible addition of quasiparticles cannot be ignored as the voltage $V_{CM}$ is swept. This potentially makes the interpretation of the period $\Delta V_{CM}$ more complicated.

\section{Estimation of $K_{IL}$ for a compressible puddle}

Here we justify the estimate $K_{IL} \lesssim \frac{1}{2} \frac{e^2}{C^g_{bulk}}$, where $C^g_{bulk} = \frac{\epsilon_z A_I}{d/2}$ is the geometric capacitance of the bulk. The upper bound is expected when the density of added quasiparticles is uniform and approaches right up to the edge of the interferometer; if the puddle is confined inwards, $K_{IL}$ will be decreased exponentially due to screening of the bulk-edge interaction by the gates, as analyzed below.

Recall the couplings $K$ are defined through the phenomenological charging energy $E = \delta n_L^2  K_L/2 + \delta n_L \delta n_I K_{IL} + \delta n_I^2  K_I/2$, where $\delta n_{I/L}$ is the charge added to the edge / bulk. 
$K_{IL}$ is then determined by the following: when charge density $\delta n_L / A_I$ is added to the bulk, what is the resulting potential produced at the edge?
When the radius of curvature is large compared to $d$, we can treat the added charge as an infinite half-plane, with charge distribution  $\delta n(x, y) = \theta(x) e  \delta n_L / A_I$.
Deep in the bulk, $x \gg 0$, this distribution produces potential $\phi = e \delta n_L /  (A_I c_g)$, where $c_g = \epsilon_z / (d / 2)$ is the capacitance per unit area to the double gates. On the other hand, for $x \ll 0$ $\phi = 0$. By reflection symmetry,  $\phi = e \delta n_L / 2 A_I c_g$ at $x = 0$.
For an edge at $x=0$, we thus conclude $K_{IL}  = e \phi(x=0) / \delta n_L \approx e^2 / 2 C^g_{bulk}$.

To determine the correction when the puddle is displaced inwards, we can solve for the full spatial dependence $\phi(x)$.
 We first solve Poisson's equation in the presence of the double gate to conclude a line charge $\rho$ at $x=0$ produces a potential 
\begin{align}
\phi(x) = -\frac{\rho}{2 \pi \sqrt{\epsilon_{xy} \epsilon_z}}\log(\tanh(\pi x \sqrt{\epsilon_{z} / \epsilon_{xy} } / 4 d))
\end{align}
where $d$ is the gate distance and $\epsilon_{xy, z}$ is the anisotropic hBN dielectric constant. 
The total potential produced at the $x=0$ edge by the $x > w$ puddle of density $\delta n_L / A_I$ is thus
\begin{align}
\phi_{IL} =  -e \frac{\delta n_L / A_I }{2 \pi \sqrt{\epsilon_{xy} \epsilon_z}} \int_{w}^\infty dy\log(\tanh(\pi y \sqrt{ \epsilon_z/ \epsilon_{xy}  } / 4 d))
\end{align}

The geometric estimate is thus $K_{IL} = e \phi_{IL} / \delta n_L$, from which  it follows that
\begin{align}
K_{IL} / (e^2 / A_I c_g) &= - \frac{\frac{1}{2 \pi \sqrt{\epsilon_{xy} \epsilon_z}} \int_{w}^\infty dy\log(\tanh(\pi y \sqrt{ \epsilon_z / \epsilon_{xy} } / 4 d))}{\frac{d}{2 \epsilon_z}} \\
& =  - \frac{4 }{ \pi^2} \int_{ \pi \sqrt{ \epsilon_z/\epsilon_{xy}} w/ 4 d}^\infty ds \log(\tanh( s ))
\end{align}
When $w \to 0$, we obtain the limit $K_{IL} / (e^2 / A_I c_g) \to 1/2$. For $w  \gg d$, the bulk edge coupling falls of exponentially due to the screening from the gates, and asymptotically we find
\begin{align}
\lim_{w \gg d} K_{IL} / (e^2 / C^g_{bulk}) \to \frac{4}{\pi^2} e^{-\pi \sqrt{ \epsilon_z/\epsilon_{xy}} w / 2  d}
\end{align}

Taking $\epsilon_z = 3.25, \epsilon_{xy} = 6.6$, and $ (e^2 / A_I c_g) / K_{IL} = 5.8$, we estimate $w / d =  0.78$, so $w = \SI{35}{nm}$.\\

%If we want to investigate screening effects arising from density-of-states corrections to the bulk capacitance (typically referred to as quantum capacitance), we can take a Thomas-Fermi approximation and assume charges are exponentially screened with a characteristic length, $k_0 = \sqrt{4\pi e^2 \frac{\mathrm d n}{\mathrm d \mu}}$, set by the inverse-compressibility.  This allows us to re-estimate $K_{IL}$ in the $w \rightarrow 0$ limit such that:

%\begin{equation}
%    K_{IL} / (e^2 / A_I c_g) = - \frac{1 }{\pi} \int_{ 0}^{\frac{\sqrt{\epsilon_{z}/ \epsilon_{xy}}}{k_0d}} ds \log(\tanh(\pi s  / 4 ))
%\end{equation}

%This integral converges and goes to zero in the limit where $k_0 d \rightarrow \infty$, i.e., in the limit of 

\section{Extraction of the quasiparticle switching rate}

We find that the timescale of the switching noise, which we interpret as the quasiparticle relaxation time $\tau$, can vary over a wide range of scales, leading to the qualitative distinction between soft and hard phase slips discussed in the main text. In Fig.~2 a-c, this timescale goes from less than \SI{10}{ms} (the time to acquire a single pixel in the measurement) to greater than \SI{7}{s} (the time to acquire a complete line trace). This dramatic increase happens over a range of only \SI{500}{mT}. In the range where $\tau \approx \SI{1}{s}$, it is experimentally convenient to extract the timescale precisely and determine quantitatively the dependence on magnetic field in this regime. Here we describe the process used to extract $\tau$ from the interference data in this ``switchy'' crossover regime from soft to hard phase slips, at the field points plotted in Fig. ~2f.

\begin{figure}[h!]
    \centering
    \includegraphics[width = 184mm]{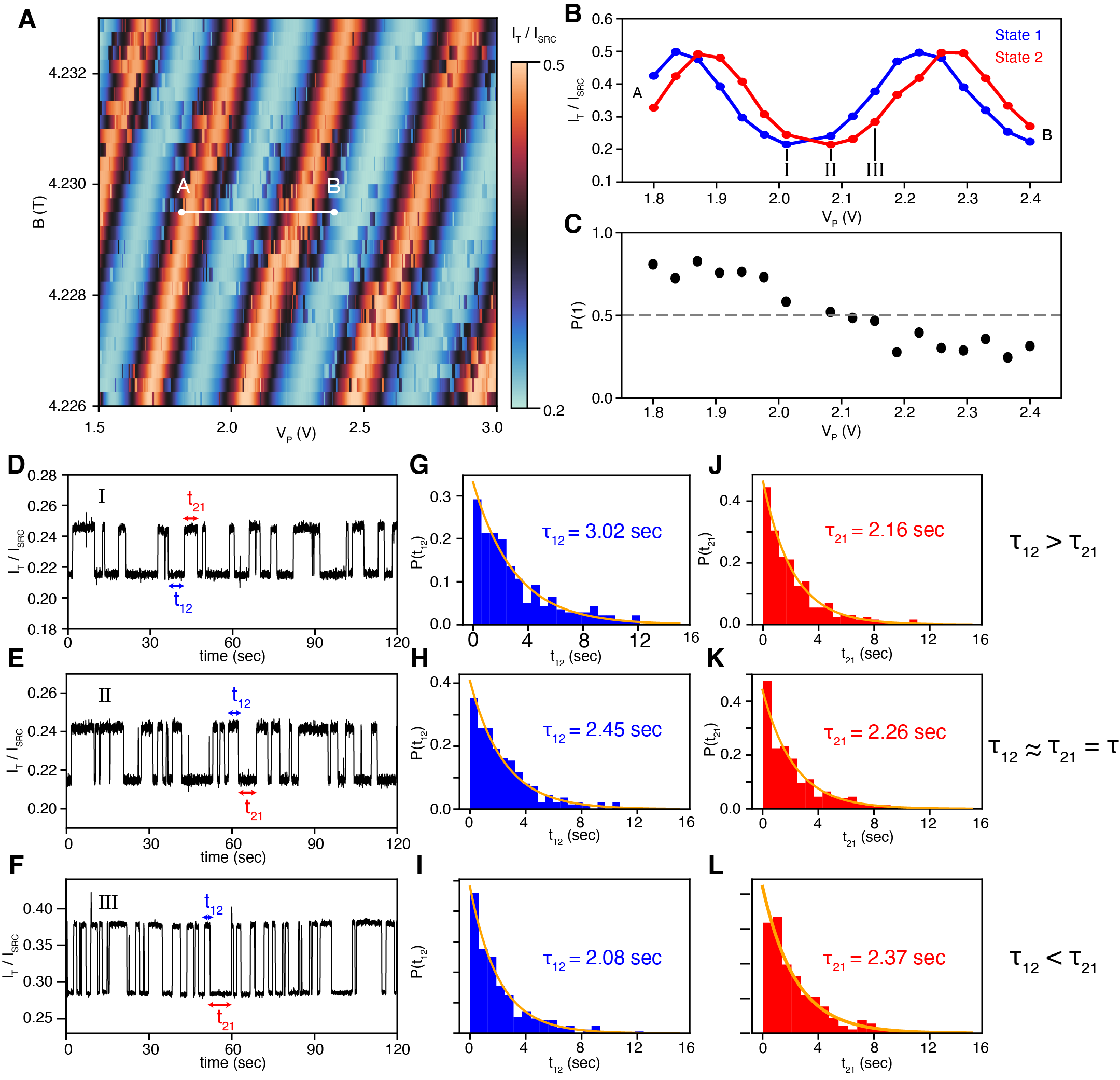}
    \caption{\textbf{Analysis of the quasiparticle switching rate in the switching-noise regime in $\nu = -1$} \textbf{(A)} Measurement of the interference pattern in the transmitted current across a single charge degeneracy line in the switching noise regime of $\nu = -1$. \textbf{(B)} The median current in each of the two stable configurations in the switching noise regime. The median value of $I_T$ in each state, plotted in red and blue, are extracted from 20 minute measurements of the current at each fixed $V_P$, by splitting the histogram of the acquired data into two halves at the midpoint and taking the median value of current for each half. \textbf{(C)} The fractional probability to occupy state 1 at each plunger gate voltage in B, acquired by simply taking the number of data points lying on either side of the halfway-point threshold between the two stable values of current. The charge degeneracy point is defined as the point at which P(1) crosses 0.5. \textbf{(D,E,F)} Three examples of the time-dependent current data used to extract the dwell times $\tau_{12}$ and $\tau_{21}$. \textbf{G,H,I} Histograms of the dwell times in state 1, $t_{12}$, each showing an exponential distribution with a characteristic time $\tau_{12}$ that decreases as the charge degeneracy line is crossed from left to right. \textbf{(J,K,L)} Histograms of the dwell times in state 2, $t_{21}$, showing that the characteristic time $\tau_{21}$ \textit{increases} as the charge degeneracy line is crossed from left to right. }
    \label{fig:supp_integer_tau_analysis}
\end{figure}

We analyze the telegraph noise using a two-state switching model, where our readout is the transmitted current, $I_T$, across the interferometer at distinct quasiparticle charge-degenerate lines for several values of magnetic field. Fig. \ref{fig:supp_integer_tau_analysis}a shows a measurement of $I_T$ illustrating one such quasiparticle charge-degenerate lines. Precisely at a charge-degeneracy point, we expect the tunneling rate to have a single characteristic time $\tau$.  In general, however, the ratio $P_{1 \rightarrow 2} / P_{2 \rightarrow 1} = e^{-\Delta / k_b T}$, where $P_{1 \rightarrow 2}$ is the probability for an electron to hop from state 1 to state 2, $P_{2 \rightarrow 1}$ is the probability for the electron to hop from state 2 to 1, and $\Delta$ is the detuning away from the charge degeneracy point in energy. As a result, the location of the charge degeneracy point in $V_P-B$ plane must be determined systematically by measuring the switching noise as a function of this detuning by sweeping the plunger gate, and only at the charge degenerate point exactly will we obtain a single timescale, which we will define as $\tau \equiv \tau_{12} \approx \tau_{21}$.

Fig. \ref{fig:supp_integer_tau_analysis}B shows the characteristic values of the transmitted current $I_T$ between which the signal switches, extracted from 20- minute measurements of the transmitted current while fixed at each point, along the illustrated line from A to B in Fig. \ref{fig:supp_integer_tau_analysis}A. The charge degenerate point is determined by plotting the fraction of time spent in state 1, $P(1)$, as function of the plunger gate voltage, plotted in Fig. \ref{fig:supp_integer_tau_analysis}C.

Fig. \ref{fig:supp_integer_tau_analysis}G-L show histograms of the two dwell times, $\tau_1$ and $\tau_2$, for different plunger gate voltages near a phase slip line, showing a strong dependence of the relative rates on detuning from charge degeneracy.  For a given $B$, we identify the the value of $V_P$ where $\tau_{12} \approx \tau_{21}$ as the charge degeneracy point, and plot $\tau\equiv (\tau_{12}+\tau_{21})/2$ in Fig. 2f, with the difference $\Delta \tau \equiv |\tau_{12} - \tau_{21}|$ as the error bar. We find that $\tau$ increases by nearly one order of magnitude over a $\SI{45}{mT}$ range of magnetic fields.